\documentclass[notitlepage, superscriptaddress, showpacs, nobalancelastpage, twocolumn, aps, prl, 10pt, floatfix]{revtex4-2}
\usepackage[english]{babel}
\RequirePackage[T1]{fontenc}
\RequirePackage{times} 

\usepackage{siunitx} 
\usepackage[italicdiff]{physics}
\usepackage{amsfonts}
\usepackage{subfigure}
\usepackage{amsmath}
\usepackage{amssymb}
\usepackage{graphicx, epstopdf}
\usepackage{appendix}
\usepackage{xspace}
\usepackage{lipsum}
\usepackage{array}
\usepackage{hyperref}
\usepackage[dvipsnames]{xcolor}
\usepackage{my_symbols}
\usepackage{CJK}
\usepackage{bm}
\usepackage{tikz}

 \newcommand{\Fig}[1]{Fig. \ref{#1}}

\begin{document}
\begin{CJK*}{UTF8}{gbsn}
	\title{Spin wave driven domain wall motion in easy-plane ferromagnets: a particle perspective}

	\author{Jin Lan (兰金)}
	\email[Corresponding author:~]{lanjin@tju.edu.cn}
	\affiliation{Center for Joint Quantum Studies and Department of Physics, School of Science, Tianjin University, 92 Weijin Road, Tianjin 300072, China}

	\author{Jiang Xiao (萧江)}
	\email[Corresponding author:~]{xiaojiang@fudan.edu.cn}
	\affiliation{Department of Physics and State Key Laboratory of Surface Physics, Fudan University, Shanghai 200433, China}
	\affiliation{Institute for Nanoelectronics Devices and Quantum Computing, Fudan University, Shanghai 200433, China}
	\affiliation{Shanghai Research Center for Quantum Sciences, Shanghai 201315, China}
	\affiliation{Shanghai Qi Zhi Institute, 200232 Shanghai, China}
	\affiliation{Zhangjiang Fudan International Innovation Center, Fudan University, Shanghai 201210, China}

	\begin{abstract}
		In easy-plane ferromagnets,
		we show that the interplay between a domain wall and a spin wave packet can be formulated as the collision of two massive particles with a gravity-like attraction.
		In the presence of magnetic dissipation, the domain wall mimics a particle subject to viscous friction, while the spin wave packet resembles a particle of variable mass.
		Due to attractive nature of the interaction, the domain wall acquires a backward displacement as a spin wave packet penetrating the domain wall, even though there is no change in  momentum of the wave packet before and after penetration.
	\end{abstract}
	\maketitle
\end{CJK*}

\emph{Introduction.}
Magnetic domain wall motion is widely used in manipulating the magnetic information for both storage and processing \cite{parkin_Magnetic_2008, parkin_Memory_2015, yu_Magnetic_2020}, and its understanding is crucial for both fundamental physics and technological applications \cite{allwood_Magnetic_2005, luo_Chirally_2019, luo_Currentdriven_2020}.
Typical approaches to drive the domain wall motion include the magnetic field \cite{schryer_Motion_1974a, thiaville_Domain_2002, nakatani_Faster_2003}, current-induced spin-transfer torque \cite{berger_Low_1978, yamaguchi_RealSpace_2004, tatara_Theory_2004, thiaville_Domain_2004} and spin-orbit torque \cite{miron_Fast_2011a, emori_Currentdriven_2013, ryu_Chiral_2013}, as well as spin waves \cite{yan_AllMagnonic_2011, wang_Domain_2012, wang_magnondriven_2015, kim_Propulsion_2014a, tveten_Antiferromagnetic_2014, qaiumzadeh_Controlling_2017, yu_Polarizationselective_2018,oh_Bidirectional_2019a,  rodrigues_SpinWave_2021,han_Mutual_2019}.
Due to the intrinsic magnetic nature, the spin wave driven domain wall motion is of special interest toward purely magnetic computing \cite{ han_Mutual_2019, lan_SpinWave_2015a,yu_Magnetic_2021}.

Investigations on the interplay between spin wave and domain wall are complicated by the fast magnetization oscillations of spin wave in both time and space, as well as the inhomogeneous magnetization of domain wall.
To overcome this complexity, a common and powerful approach is to make use of the linear or angular momentum conservation \cite{yu_Magnetic_2021}, which focuses on global momentum transfer and avoids local interaction details.

In easy-axis ferromagnets, the spin wave can either drag or push the domain wall, depending on whether the spin wave is transmitted or reflected \cite{yan_AllMagnonic_2011, wang_Domain_2012, wang_magnondriven_2015}.
And when extending to antiferromagnetic and ferrimagnetic environments, the direction of the domain wall motion can be controlled by tuning the spin wave polarization \cite{tveten_Antiferromagnetic_2014, qaiumzadeh_Controlling_2017, yu_Polarizationselective_2018, oh_Bidirectional_2019a} or frequency \cite{rodrigues_SpinWave_2021}.
Easy-plane magnet is another important category of magnetic materials, and is a fertile ground for emerging physics including magnon superfluid \cite{skarsvag_Spin_2015, flebus_TwoFluid_2016, upadhyaya_Magnetic_2017}, magnetic vortex \cite{tretiakov_Dynamics_2008, dasgupta_Quantum_2020}, bimeron \cite{lin_Skyrmion_2015, yu_Transformation_2018}.
However, domain wall in easy-plane ferromagnets is only studied in limited cases \cite{mikeska_Solitons_1977, how_Soliton_1989, kosevich_Magnetic_1990, hill_SpinTorqueBiased_2018}, and its motion driven by spin wave remains elusive.

In this work, we demonstrate that the domain wall in easy-plane ferromagnets is  displaced toward the spin wave source as  the  wave  passing through the wall, albeit spin wave carries the same momentum before and after penetration.
By developing a unified Lagrangian framework, we transform the interplay between a spin wave packet and a domain wall to an equivalent collision  process between two massive particles subject to gravity-like attraction, which greatly simplifies  the wave-soliton interaction scenario and enables a classical yet intuitive understanding.

\begin{figure}[t]
	\centering
	\includegraphics[width=0.49 \textwidth, trim=30 0 30 30, clip]{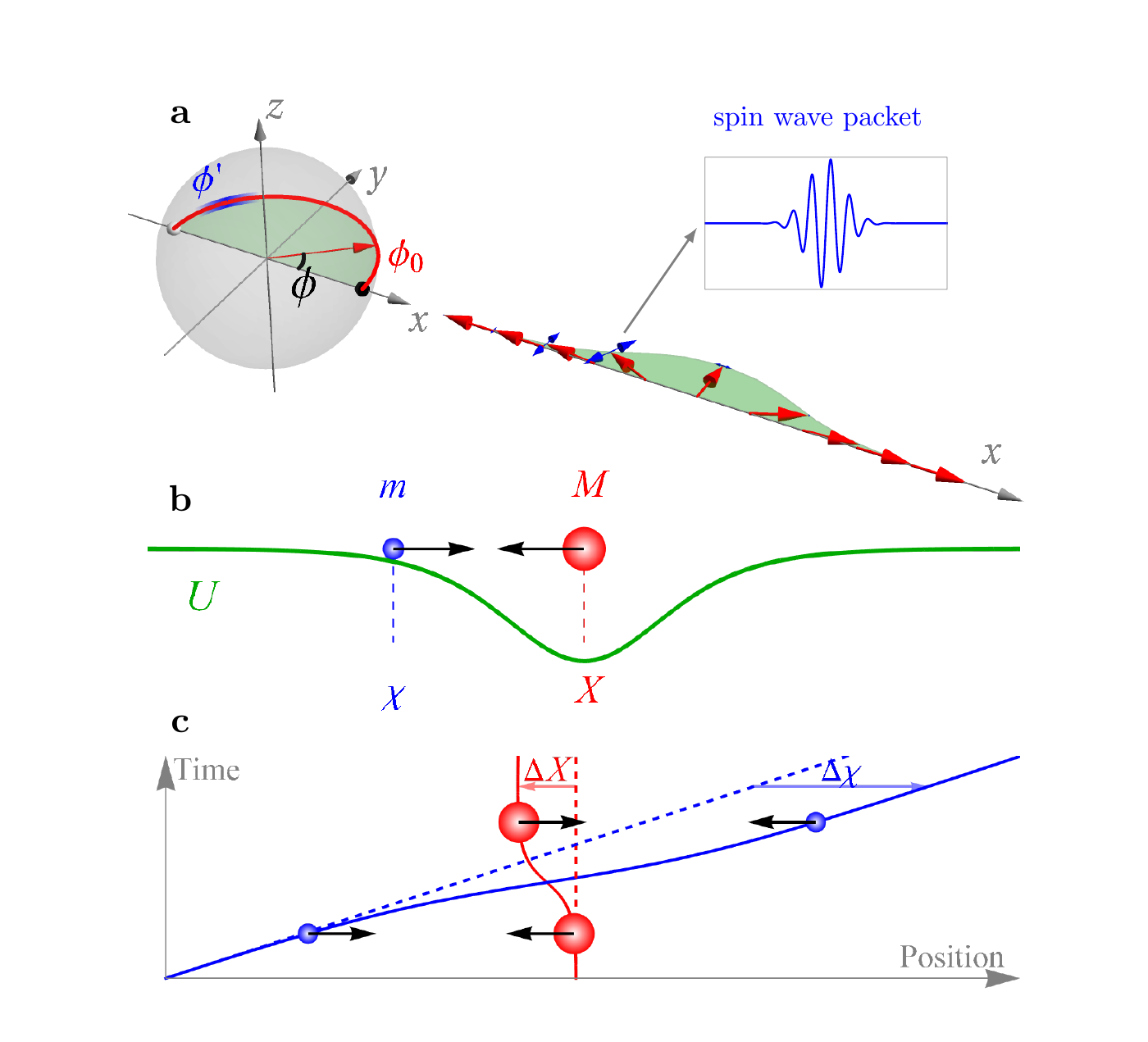}
	\caption{{\bf Schematics of the interplay between a spin wave packet and a magnetic domain wall in easy-plane ferromagnets.}
		(a) Schematics of the magnetic profile.
		The red (blue) arrows depict the magnetizations of a domain wall (spin wave packet), and the green plane denotes the anisotropy-defined easy plane.
		(b) Schematics of the equivalent particle model.
		The red (blue) ball denotes the domain wall (spin wave packet), and the green line plots the interaction potential $U$.
		(c) Schematics of the forward and backward jumps.
		The solid (dashed) lines plot the loci of a heavy and a light particle during penetration with (without) mutual attraction, and the overall differences of two loci indicated by $\Delta X$ and $\Delta \chi$, respectively.
		\label{fig:sch}
	}
\end{figure}

\emph{Basic model.}
We consider a one-dimensional ferromagnetic wire extending along $x$-axis as shown in Fig. \ref{fig:sch}(a),
where the magnetization direction (red arrows) is denoted by a unit vector $\mb(x)$. We assume that the ferromagnet has a strong hard-axis anisotropy along the $\hbz$-direction, and thus $x$-$y$ plane is the easy plane. In terms of the $z$-component of magnetization $m_z$, and the azimuthal angles $\phi$ with respect to $z$-axis, the magnetization direction is explicitly denoted by $\mb \equiv ( \sqrt{1-m_z^2}\cos\phi, \sqrt{1-m_z^2} \sin\phi, m_z)$.
The magnetic free energy  is  $\cE = (S/2) \int [K'm_z^2 + K\sin^2\phi + A(\partial_x\mb)^2] dx$, where $K'$ is a strong easy-plane anisotropy favoring $x$-$y$ plane, $K\ll K'$ is a weak easy-axis anisotropy along $x$-axis, and $A$ is the exchange coupling constant.
Here $S=\mu_0M_s\cA$ is the magnetic flux,  where $\mu_0$ is the vacuum permeability, $M_s$ is the saturation magnetization and $\cA$ is the cross-section area of the magnetic wire.
The main effect of dipolar field is to renormalize the anisotropy constants in exchange regime, thus is not included explicitly  in this work.

The Lagrangian of the magnetic system reads \cite{kosevich_Magnetic_1990, thiaville_Domain_2002}
\begin{equation}
	\label{eqn:Lag_mag}
	\cL = \frac{S}{\gamma} \int m_z\dot{\phi}\, dx - \cE,
\end{equation}
where the first term is the kinetic energy of magnetic system \cite{thiaville_Domain_2002}, and $\gamma$ is the gyromagnetic ratio.
In addition, the accompanying Rayleigh function accounting for the dissipation is $\cR = (\alpha S/2\gamma) \int \dot{\mb}^2 dx$, with $\alpha$ the Gilbert damping constant.
The Euler-Lagrangian variation of $\cL$ in \Eq{eqn:Lag_mag} and the accompanying $\cR$ with respect to $\mb$ yields the Landau-Lifshitz-Gilbert (LLG) equation $\dot{\mb} = - \gamma \mb \times \bh +\alpha \mb\times \dot{\mb}$, where $\bh=-S^{-1}\delta \cE/\delta \mb$ is the effective magnetic field acting on magnetization $\mb$.

Due to the strong easy-plane anisotropy $K'$, the magnetization lies predominatingly in the $x$-$y$ plane with $m_z\simeq 0$, as illustrated in Fig. \ref{fig:sch}(a).
Hence, the Lagrangian  simplifies to
\begin{align}
	\label{eqn:Lag_phi}
	\cL =\frac{S}{2} \int \qty[\frac{\dot{\phi}^2}{ \gamma^2 K'} - A (\partial_x \phi)^2 +K \sin^2\phi]dx,
\end{align}
and the Rayleigh function simplifies to $\cR = (\alpha S/2\gamma)\int \dot{\phi}^2 dx$.
Correspondingly, the magnetic dynamics reduces to a damped sine-Gordon equation on the azimuthal angle $\phi$ \cite{mikeska_Solitons_1977, how_Soliton_1989, kosevich_Magnetic_1990, hill_SpinTorqueBiased_2018, caretta_Relativistic_2020}
\begin{equation}
	\label{eqn:SG}
	-\frac{1}{c^2} \ddot{\phi} -\frac{\beta}{c^2} \dot{\phi} = - \partial_x^2\phi + \frac{ \sin 2\phi}{2 W^2},
\end{equation}
where $c=\gamma \sqrt{AK'}$ is the effective `speed-of-light' of the magnetic system, $W=\sqrt{A/K}$ is the characteristic magnetic length (or the domain wall width), and $\beta =\alpha\gamma K'$ is the dissipation coefficient.

\emph{Domain wall and spin wave packet.}
The dynamics of the azimuthal angle $\phi(x, t)$ in \Eq{eqn:SG} can be naturally divided into the slow dynamics due to the domain wall motion and the fast dynamics due to the spin wave excitation, \emph{i.e.}, $\phi = \phi_0 + \phi'$.
In the following, we first investigate the dynamics of $\phi_0$ and $\phi'$ separately, and then their interaction caused by the nonlinearity embedded in the $\sin 2\phi$ term of \Eq{eqn:SG}.

It is well known that the static domain wall solution hosted by Eq. \eqref{eqn:SG} has the following  soliton form \cite{Vachaspati_2006zz, cuevas-maraver_SineGordon_2014}
\begin{align}
	\label{eqn:dw_phi}
	\phi_0(x, t) = 2\arctan\qty[\exp(- \frac{x-X(t)}{W} )],
\end{align}
where $X$ is the domain wall central position.
The domain wall profile in \Eq{eqn:dw_phi} corresponds to the magnetization rotating steadily from $\mb = -\hbx$ ($\phi_0=\pi$) at $x\ll X$ to $\mb = +\hbx$ ($\phi_0=0$) at $x\gg X$, as illustrated in Fig. \ref{fig:sch}(a).
Due to invariance of the topological charge \cite{Vachaspati_2006zz, cuevas-maraver_SineGordon_2014}: $Q=(1/\pi)\int d\phi = -1$,
the moving domain wall in \Eq{eqn:dw_phi} maintains a relatively fixed shape, hence its evolution is  mainly determined  by the variation of its central position $X(t)$: $\phi_0(t)\equiv \phi_0[X(t)]$.

In the meantime, we consider a spin wave packet in the Gaussian form
\begin{equation}
	\label{eqn:sw_phi}
	\phi'(x, t) =\frac{1 }{k_0W}\sqrt{\frac{2 n\hbar\omega_0}{\sqrt{\pi}\sigma SK }}
	e^{-\frac{[x-\chi(t)]^2}{2\sigma^2} }
	\cos\qty{ k_0 [x-\chi(t)]},\\
\end{equation}
where $n$ is the magnon number representing the spin wave intensity, $\sigma$ is the typical width, $\chi$ is the central position,  $k_0$ and $\omega_0$ are the central wavevector and frequency of the wave packet.
For exchange-type spin wave under investigation in this work, we may  assume $k_0 W\gg 1$, then the spin wave packet is narrow  in both spatial and wavevector spaces, i.e.  $\sigma \ll W$ and $1/\sigma\ll k_0$ \cite{lan_Skew_2021}.
Hence, the evolution of the spin wave packet can also be described by the variation of its central position $\chi(t)$:  $\phi'(t)\equiv \phi'[\chi(t)]$.

\emph{Particle collision model.}
As seen from \Eq{eqn:dw_phi} and \Eq{eqn:sw_phi}, both the soliton-like domain wall and the spin wave packet are reduced to particle-like objects characterized by their positions: $X(t)$ for the domain wall and $\chi(t)$ for the spin wave packet.
In terms of these two degrees of freedom, the Lagrangian is recast from \Eq{eqn:Lag_phi} to
\begin{equation}
	\label{eqn:Lag}
	\cL = \frac{M}{2}\dot{X}^2+ \frac{m}{2}\dot{\chi}^2-U,
\end{equation}
where the effective masses of the domain wall and spin wave packet are defined by their static energies as $Mc^2=2SKW$ and $mc^2=n\hbar \omega_0$, respectively.
The interaction energy $U$ in \Eq{eqn:Lag}, originated from the $\sin2\phi$ term in the sine-Gordon equation, takes the following attractive gravity-like form
\begin{equation}
	\label{eqn:U_int}
	U = -GMm\frac{\sech^2\qty[(\chi-X)/W]}{2W},
\end{equation}
where $G= c^4/( SA k_0^2) $ is the effective gravitational constant.
This interaction energy $U$ is collaboratively caused by the inhomogeneous domain wall magnetization and the reduction of its magnitude by spin wave \cite{lan_Skew_2021}.
It is noteworthy that the static energies of the domain wall and spin wave packet $Mc^2$ and $mc^2$ are omitted in the Lagrangian of \Eq{eqn:Lag} since they are both constants.
\Eq{eqn:Lag} and \Eq{eqn:U_int} indicate that the interplay between a domain wall and a spin wave packet in easy-plane ferromagnets can be viewed as the collision of two particles with mass $m$ and $M$ subject to a gravitational-like attraction of energy $U$, as illustrated in Fig. \ref{fig:sch}(b).

Similar to the transformed Lagrangian in \Eq{eqn:Lag}, the Rayleigh dissipation function is transformed to $\cR = \beta M \dot{X}^2/2+ \beta m \dot{\chi}^2/2$,
where the dissipation of domain wall and spin wave packet share the same coefficient $\beta$.
Because of the topological protection, the domain wall profile in \Eq{eqn:dw_phi} maintains the same form, regardless of dissipation.
However, the intensity of spin wave packet in \Eq{eqn:sw_phi} is expected to decay  due to  dissipation.
As a result, the domain wall mass $M$ is a constant of time, but the mass $m$ of the spin wave packet reduces in the magnetic background, just like an icy ball dissolving in water.

From the Lagrangian in \Eq{eqn:Lag} and the accompanying Rayleigh dissipation function, the dynamics of the domain wall and spin wave packet are then governed by
\begin{subequations}
	\label{eqn:dw_eom_Xxm}
	\begin{align}
		\label{eqn:dw_eom_X}
		{\dv{P}{t}}: & \quad    M \ddot{X}   = -\pdv{U}{X} -\beta M \dot{X}, \\
		\label{eqn:sw_eom_x}
		{\dv{p}{t}}: & \quad
		\begin{cases}
			m \ddot{\chi} & = -\pdv{U}{\chi} \\
			\dot{m}       & = -\beta m
		\end{cases},
	\end{align}
\end{subequations}
where \Eqs{eqn:dw_eom_X}{eqn:sw_eom_x} describe the evolution of the domain wall momentum $P \equiv M\dot{X}$ and the momentum of the spin wave packet $p \equiv m\dot{\chi}$, respectively.
In the absence of dissipation ($\beta=0$) we have $d(P+p)/dt = 0$ in this isolated two-body system, \ie the total momentum is conserved or the momenta of domain wall and spin wave are exchanged to one another via mutual potential $U$.
The dissipation has different effects on the domain wall and the spin wave packet: the domain wall experiences a viscous force and slows down due to dissipation, but the spin wave packet  loses its mass by dissolving into background while maintaining its speed.
Above viscosity and dissolution scenarios represent two opposite limits of particle dynamics in a fluid, with the former and latter denoting the full resistance and compliance of mutual deformation, respectively \cite{smith_Remington_2015}.
With \Eq{eqn:dw_eom_Xxm}, we successfully transformed the highly non-trivial interaction between an inhomogeneous magnetic domain wall and a fast-oscillating spin wave packet into a simple collision scenario between a particle of constant mass $M$ and a particle with variable mass $m$.

\emph{Domain wall motion driven by spin wave packet.}
We now consider the simplest case  without dissipation ($\beta=0$),  for which \Eq{eqn:dw_eom_Xxm} becomes $M\ddot{X} = -m\ddot{\chi}=-\partial_X U$.
Due to the attractive nature of $U$, as the spin wave packet passing through the domain wall,  the wave packet and the domain wall experience a forward and backward jump, respectively, as denoted by $\Delta X$ and $\Delta\chi$ in Fig. \ref{fig:sch}(c).

More explicitly, a spin wave packet is typically  much  lighter than the domain wall: $m\ll M$, therefore we may regard the domain wall more or less static as the packet passing through the domain wall. The wave packet velocity at position $\chi$ can be approximately evaluated via the energy conservation $ mv_\chi^2/2- m v_0^2/2=-U(\chi-X)$, which yields
\begin{equation}
	\label{eqn:sw_dv}
	v_\chi - v_0 \simeq -{\frac{U(\chi-X)}{ mv_0}}  > 0,
\end{equation}
where $v_0 = v_{\chi\ra \pm\infty}$  is the initial and final velocity of the packet before and after penetration with $U = 0$.
Because of this velocity enhancement, the spin wave packet gains a velocity inside the domain wall, thus leading to a forward jump in comparison to the case  without  the domain wall:
\begin{equation}
	\label{eqn:sw_dx}
	\Delta\chi = \int \qty(v_\chi - v_0) dt
	\simeq -\int \frac{U}{ mv_0^2} d\chi
	= \frac{GM}{v_0^2}.
\end{equation}
In turn, the domain wall acquires a backward jump due to momentum conservation:
\begin{equation}
	\label{eqn:dw_dX}
	\Delta X = -\frac{m}{M}\Delta\chi = -\frac{Gm}{v_0^2}.
\end{equation}
Here, we focus on the displacement of the domain wall, because the domain wall stops once the wave packet leaves the domain wall behind.

In the presence of dissipation ($\beta\neq 0$), the dynamics of packet position $\chi$ is unaltered in \Eq{eqn:sw_eom_x}, hence the packet forward jump $\Delta\chi$ in \Eq{eqn:sw_dx} remains the same.
However, the domain wall velocity $V$ and the wave packet mass $m$ are subjected to a similar form of dissipation in $-\beta V$ and $-\beta m$, which means that both of them decrease by a common factor of $e^{-\beta t}$.
Consequently, the domain wall displacement in \Eq{eqn:dw_dX} is modified to
\begin{align}
	\label{eqn:dw_dX_beta}
	\Delta X = -\frac{ Gm_0 }{v^2_0} \exp(\beta \frac{\chi_0-X_0}{v_0}),
\end{align}
where $m_0$ and $\chi_0$ are the initial mass and position of the spin wave packet, and $X_0$ is the initial position of domain wall.

\begin{figure}[bt]
	\centering
	\includegraphics[width=0.49\textwidth]{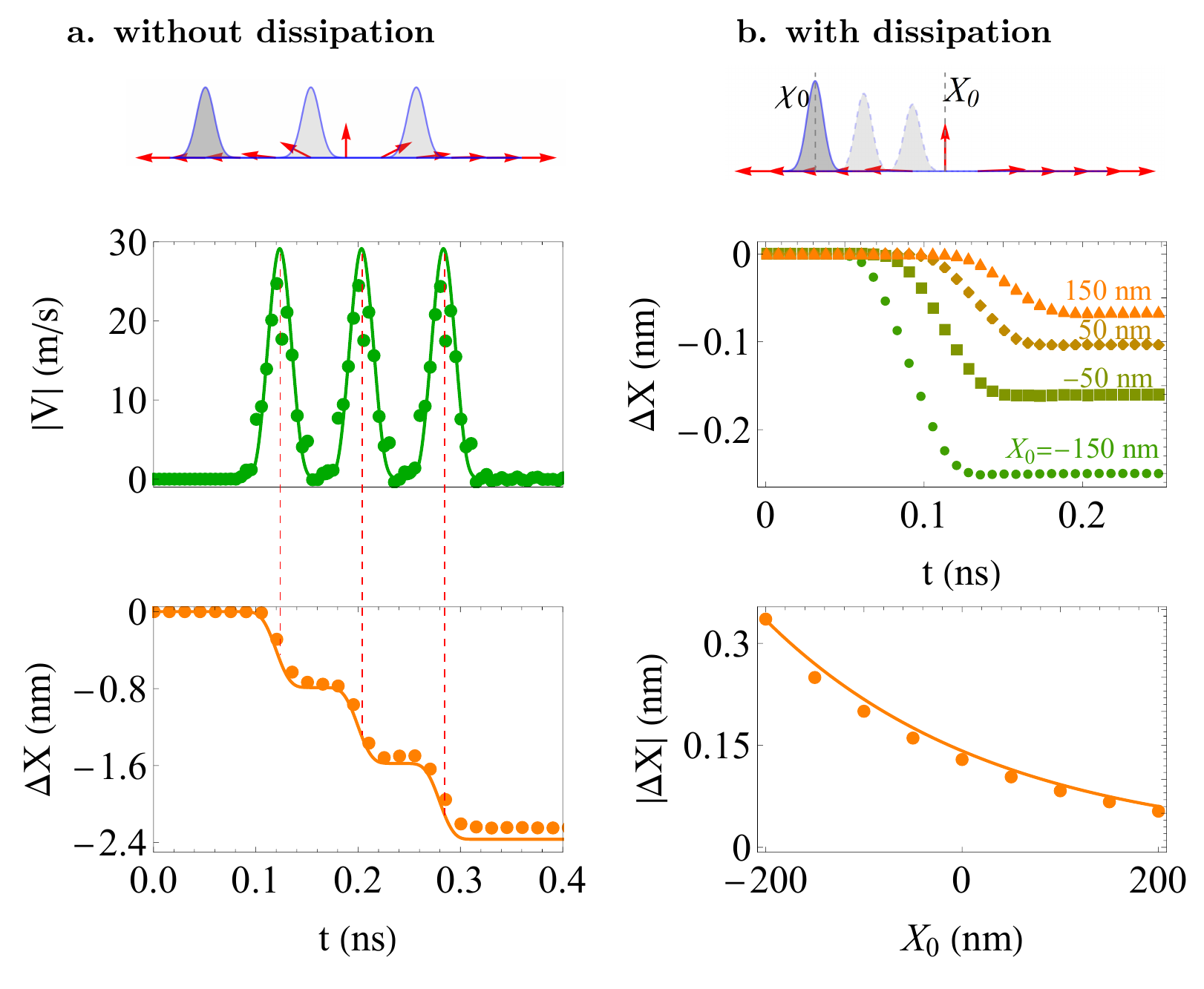}
	\caption{{\bf Domain wall motion induced by spin wave packets in a magnetic wire with (a) without (b) with dissipation.}
		In (a), the upper panel is the schematics, the middle and lower panels plot evolution of velocity $V$ and displacement $\Delta X$ induced by three consecutive packets, respectively.
		In (b), the upper panel is the schematics, the middle and lower panels plot displacement $\Delta X$ as function of time $t$ and domain wall position $X_0$, respectively.
		In each plot, solid lines are calculated based on the particle-collision model, and dots are extracted from micromagnetic simulations.
		The damping constant is $\alpha=1.0\times 10^{-5}$ in (a) and $\alpha=0.006$ in (b).
		\label{fig:sw_3wp}
	}
\end{figure}

The domain wall displacements $\Delta X$ in \Eq{eqn:dw_dX} and \Eq{eqn:dw_dX_beta}, as formulated in above particle collision scenarios, are confirmed by micromagnetic simulations based on COMSOL Multiphysics \cite{COMSOL}, which are  shown in \Fig{fig:sw_3wp}.
The magnetic parameters used in simulations are: the exchange constant $A = \SI{3.28e-11}{\ampere\meter}$, the easy-axis anisotropy $K = \SI{3.88e4}{\ampere\meter}$, the easy-plane anisotropy $K' = \SI{1.552e7}{\ampere\per\meter}$, the gyroscopic ratio $\gamma = \SI{2.21e5}{\meter\per\ampere\per\second}$, the Gilbert damping constant $\alpha=0.006$.
The spin wave packet is excited with a central frequency of $f=\SI{100}{GHz}$, or a normalized wavevector $k_0 W\approx 3.5$.
The corresponding parameters in particle collision model are: the domain wall mass $ M = \SI{2.2e-27}{kg}$, the spin wave packet mass $m = \SI{3.6e-28}{kg}$,
the gravitational constant $G = \SI{5.3e25}{m^3.kg^{-1}.s^{-2}}$, dissipation coefficient $\beta = \SI{2.1e10}{s^{-1}} $, and the initial packet velocity $v_0 \approx \SI{4.8e3}{m.s^{-1}} $.
As three successive spin wave packets penetrate, the domain wall acquires three velocity pulses, and experiences the same amount of negative displacement for each packet in \Fig{fig:sw_3wp}(a).
And when dissipation becomes remarkable in \Fig{fig:sw_3wp}(b), the domain wall residing at further positions away from the spin wave source is subject to smaller displacement, for which the exponential decay law in \Eq{eqn:dw_dX_beta} is obeyed.  The domain wall velocity increases as approaching the spin wave source, which is simply because the spin wave intensity is stronger near the source.

\begin{figure}[bt]
	\centering
	\includegraphics[width=0.49\textwidth ]{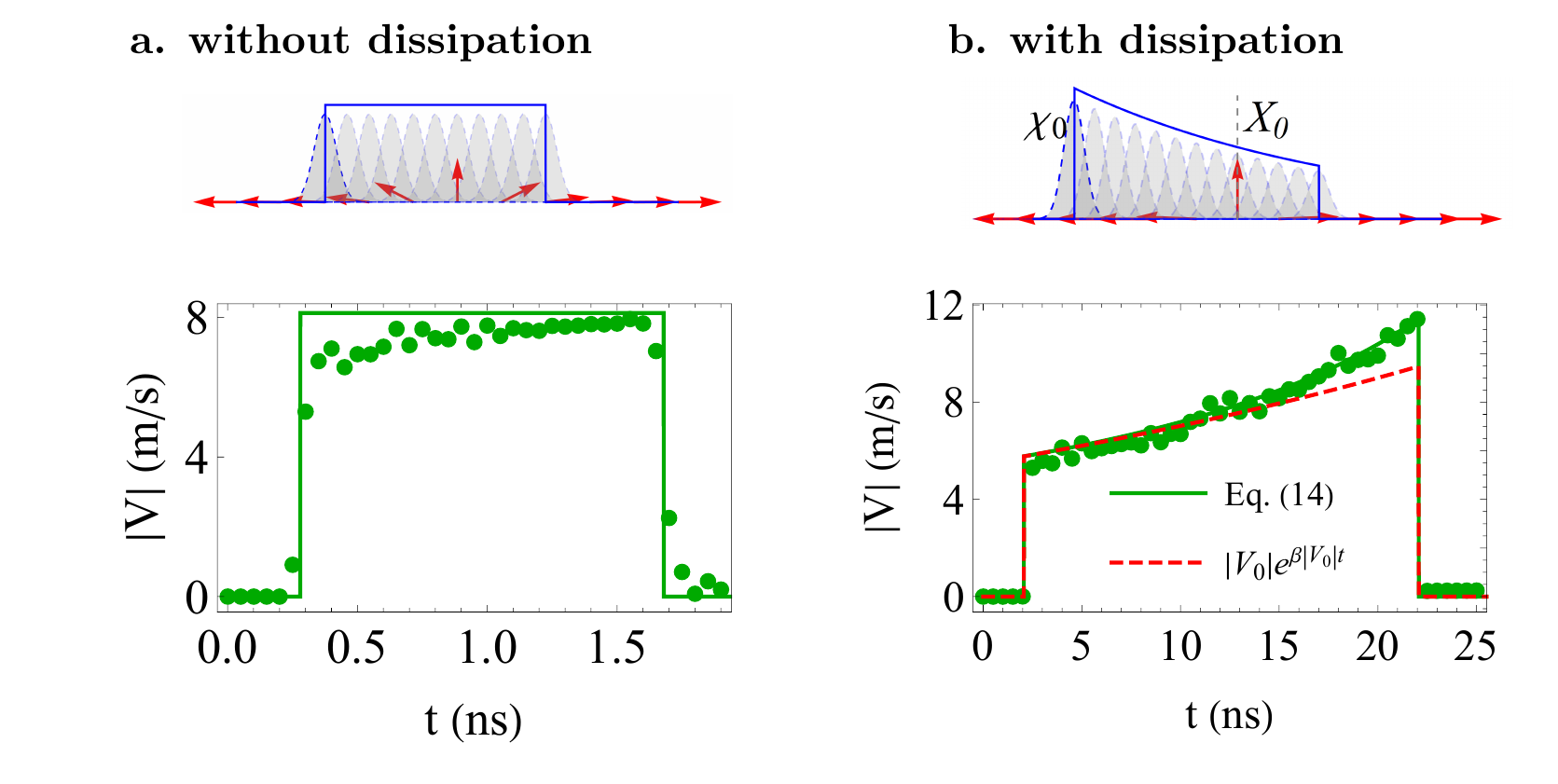}
	\caption{
		{\bf Domain wall motion induced by continuous spin wave in a magnetic wire (a) without (b) with dissipation.}
		In both (a)(b), the upper panels show the schematics, and the lower panels plot evolution of velocity $V$.
		The dots are extracted from micromagnetic simulations, the solid lines are theoretical calculations based on particle collision model, and the dashed line in (b) is based on an empirical anticipation.
		All settings and parameters follow Fig. \ref{fig:sw_3wp}.
		\label{fig:sw_cont}
	}
\end{figure}

\emph{Domain wall motion driven by continuous spin wave.}
In most experimental and theoretical considerations, continuous spin wave is excited rather than a wave packet. For such a case, we may treat the continuous spin wave as a train of spin wave packets, as schematically shown in Fig. \ref{fig:sw_cont}.
When the dissipation is absent $\beta=0$, for continuous spin wave of density $\rho$, the mass penetrating through the domain wall in a time interval $\Delta t$ is $m=\rho v_0\Delta t$.
Then according to \Eq{eqn:dw_dX}, the domain wall acquires a backward velocity
\begin{align}
	\label{eqn:dw_V}
	V  = \frac{\Delta X}{\Delta t}= - \frac{G\rho}{v_0},
\end{align}
as driven collaboratively by multiple spin wave packets within domain wall.

In the case of finite dissipation $\beta\neq 0$, if the spin wave density is $\rho_0$ at the source at $x = \chi_0$, then the spin wave density at the domain wall center $X(t)$ attenuates exponentially due to dissipation: $\rho = \rho_0\exp\qty{\beta\abs{\chi_0-X(t)}/v_0}$, hence the spin wave density experienced by a moving domain wall satisfies $\dot{\rho}/\rho=-\beta V/v_0$.
In conjunction with \Eq{eqn:dw_V}, one has $\dot{V}/V = \dot{\rho}/\rho = -\beta V/v_0$, \emph{i.e.}, the domain wall effectively experiences a drag force toward the spin wave source with drag coefficient $\beta/v_0$.
Therefore, the evolution of domain wall velocity is explicitly described by
\begin{align}
	\label{eqn:dw_V_beta}
	V = -\frac{ |V_0|}{ 1 - \frac{\beta |V_0|}{v_0} t },
\end{align}
where $V_0= -G\rho_0 \exp[\beta (\chi_0-X_0)/v_0]$ is the initial velocity at the moment that spin wave touches the domain wall.

The validity of the domain wall velocity given by \Eqs{eqn:dw_V}{eqn:dw_V_beta} are confirmed by micromagnetic simulations as shown in \Fig{fig:sw_cont}.
Irrespective of dissipation, the domain wall stops immediately once the spin wave leaves the domain wall behind, indicating that the domain wall is only temporarily driven by the spin wave during its penetration.
In Fig. \ref{fig:sw_cont}(b), the evolution of domain wall velocity with dissipation adopts the reciprocal form in \Eq{eqn:dw_V_beta} instead of an exponential growth form, endorsing the unconventional role of dissipation in shaping the interplay between domain wall and spin wave.

\emph{Discussions.} The easy-plane ferromagnet is distinct from its easy-axis counterpart in two folds: the spin wave is linearly polarized, and thus does not carry any angular momentum; the domain wall is inertial, and tends to maintain its original velocity \cite{doering_Ueber_1948, thomas_Dynamics_2010, caretta_Relativistic_2020}.
In addition, the equal spin wave amplitudes and velocities before and after penetration indicates that the domain wall motion in this work is not induced by permanent transfer of angular momentum \cite{yan_AllMagnonic_2011} or linear momentum \cite{wang_Domain_2012, wang_magnondriven_2015}
The elevated spin wave velocity inside the domain wall, on the other hand, indicates a temporary borrowing and a later return of linear momentum between the spin wave and the domain wall.

The spin wave passes through the domain wall perfectly in this work, therefore also shares much common features with Balazs thought experiment on light passing a transparent medium \cite{balazs_EnergyMomentum_1953, barnett_Resolution_2010}.
With noticeably lower `speed-of-light' in this magnetic environment, more insights on the Abraham-Minkowski dilemma are accessible \cite{barnett_Resolution_2010}.

\emph{Conclusions.}
In conclusion, based on a particle collision scenario, we show that the domain wall is dragged backward during penetration of spin wave in easy-plane ferromagnet.
The particle-based viewpoint established in this work, provides a simple yet powerful tool to analyze the interplay between soliton and fluctuation wave in various nonlinear systems \cite{cuevas-maraver_SineGordon_2014, Vachaspati_2006zz}.

\emph{Acknowledgements.}
J.L. is grateful to Gongzheng Chen, Xinghui Feng and Pujian Mao for insightful discussions.
J.L. is supported by National Natural Science Foundation of China (Grant No. 11904260) and Natural Science Foundation of Tianjin (Grant No. 20JCQNJC02020).
J.X. is supported by Science and Technology Commission of Shanghai Municipality (Grant No. 20JC1415900) and Shanghai Municipal Science and Technology Major Project (Grant No. 2019SHZDZX01).

%
  
\end{document}